\documentclass[conference]{IEEEtran}

\ifCLASSINFOpdf

\else
\fi
\usepackage{multicol}
\usepackage{tikz}
\usepackage{tkz-graph}
\usepackage{bbm}
\usepackage{booktabs}
\usepackage{flushend}
\usepackage{listings}
\usepackage{url}
\usepackage{verbatim}
\renewcommand{\texttt}[2][black]{\textcolor{#1}{\ttfamily #2}}

\usepackage{pgfplots}

\usepackage{algorithmic}

\usepackage{color}
\usepackage[cmex10]{amsmath}
\usepackage{amssymb}
\usepackage{dblfloatfix}

\usepackage{diagbox}
\begin{document}
\pagenumbering{gobble}

%
\title{\textbf{\Large JXES:  JSON Support for the XES Event Log Standard\\[-1.5ex]}}

\author{
\IEEEauthorblockN{~\\[-0.4ex]\large Madhavi Bangalore Shankara Narayana, Hossameldin Khalifa, Wil van der Aalst \\[0.3ex]\normalsize}
\IEEEauthorblockA{Process and Data Science Department, RWTH Aachen University\\
Process and Data Science department, Lehrstuhl fur Informatik 9 52074 Aachen, Germany\\
Emails: {madhavi.shankar@pads.rwth-aachen.de}, {hossameldin.khalifa@rwth-aachen.de}, {wvdaalst@pads.rwth-aachen.de}}}


%


\maketitle

\begin{abstract}
Process mining assumes the existence of an event log where each event refers to a case, an activity, and a point in time. XES is an XML based IEEE approved standard format for event logs supported by most of the process mining tools. JSON (JavaScript Object Notation) is a lightweight data-interchange format. In this paper, we present JXES, the JSON standard for the event logs and also provide implementation in ProM for importing and exporting event logs in JSON format using 4 different parsers. The evaluation results show notable performance differences between the different parsers (Simple JSON, Jackson, GSON, Jsoninter).
\end{abstract}




%
\IEEEpeerreviewmaketitle

\section{Introduction}

Process mining is a growing discipline in Data Science that aims to use event logs obtained from information systems to extract information about the business process.
A lot of research in Process mining has been about proposing techniques to discover the process models starting from event logs.

Event logs are being produced in research and practice, each system group has developed their own logging mechanism for their logging system. 
An event log can be seen as a collection of cases and a case can be seen as a trace/sequence of events. 
Different researchers and companies use different formats for storing their event logs.
Event data can come from a database system, a CSV (comma-separated values), a spreadsheet, JSON file, a transaction log, a message log and even from APIs providing data from websites or social media.

The XES (eXtensible Event Stream, an XML-based standard  has been developed \cite{XES} to support the easy exchange of event logs between different tools.
This format is supported by Celonis, Disco, UiPath, ProM, PM4Py, Apromore, QPR ProcessAnalyzer, ProcessGold, etc.
The JXES structure is based on a JSON Object, which contains all the necessary data and information for event logs.
ProM \cite{ProM} is a process mining framework that contains a large set of pre-processing, process discovery, conformance and performance/enhancement algorithms. 
In this paper, we define the JSON format for event logs and also provide new plugin implementations for importing/exporting events logs with JSON format. 
 
The reminder of this paper is structured as follows. In Section II, we discuss the related work and also define the JSON file standard for event log. In Section III, we describe the parser implementation. In Section IV we discuss the evaluation criteria. In Section V we provide information on accessing this plugin and also its demonstration information and Section VI concludes this paper.

\section{Supporting XES using JSON}
\begin{figure}[htb!]
\includegraphics[width=0.5\textwidth]{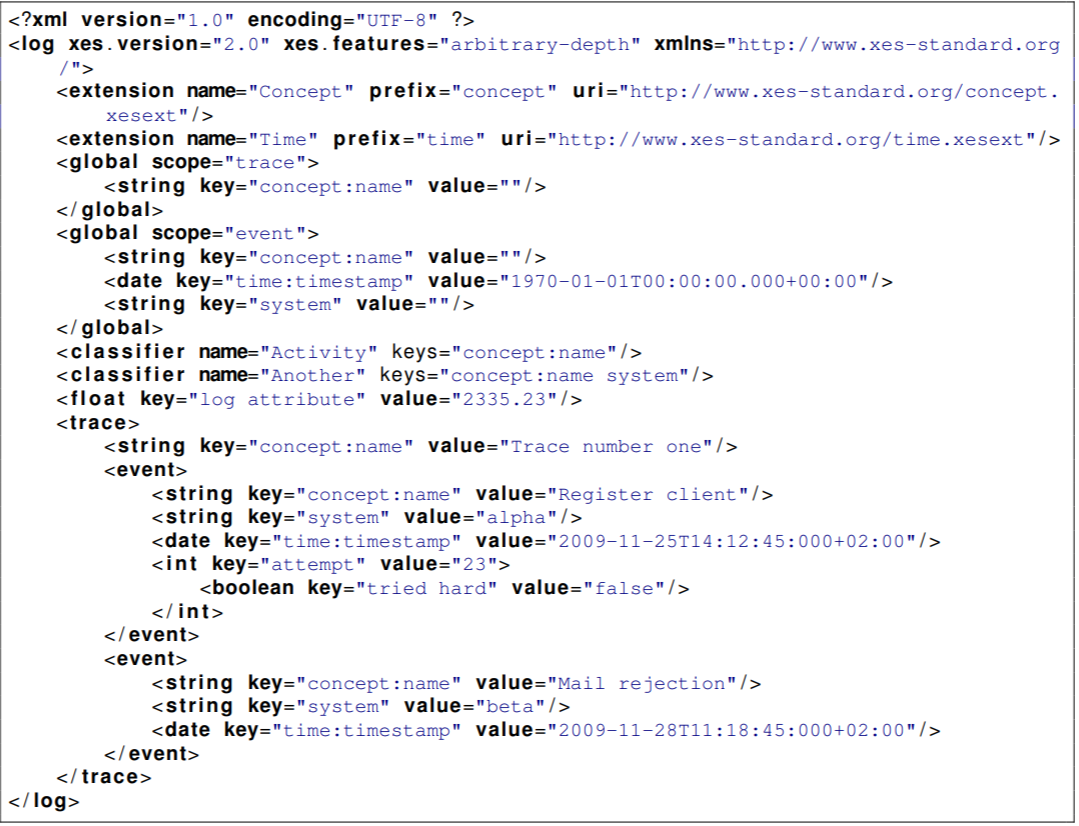}
\vspace{-0.5cm}
\caption{XES template.}
\label{fig:xesmeta}
\end{figure}

The XES standard defines a grammar for a tag-based language whose aim is to provide designers of information systems with a unified and extensible methodology for capturing systems behaviors by means of event logs and event streams. But, not all information systems support XES format and would be beneficial to extend the XES semantic to JSON format. An XML schema shown in Figure \ref{fig:xesmeta} describes the structure of an event log. 

OpenXES \cite {OpenXes} is a standard implementation of XES. OpenXES stores the event log in main memory using the standard classes provided by the Java collections framework. This simplifies the development and works well with small event logs typically used as examples for research purposes. However, when using OpenXES for large and complex real-life event logs (e.g., 10 million events with 3 attributes each), the available main memory on a typical workstation (e.g., 4 GB) is insufficient to load the event log.
XESLite \cite{xeslite} can handle large event logs and overcome the drawback of memory issue. 
DB-XES \cite{DB-XES} is a database schema which resembles
the standard XES structure.
\\
JSON is an open standard lightweight file format commonly used for data interchange. It uses human-readable text to store and transmit data objects.
Hence, the motivation to define a JSON event log format and create a plugin lies in the fact that we can easily use the JSON logs. We have used the 4 design principles defined for the XES format, namely, (i) Simplicity, (ii) Flexibility, (iii) Extensibility and (iv) Expressivity. This helped us to make design decisions with respect to defining the JXES standard format as well as evaluate the implementations and make suggestions on the optimal parser 

For defining the JSON standard format, we have taken into account the XES meta-model shown in Figure \ref{fig:xesmeta} which is represented by the basic structure (log, trace and event), Attributes, Nested Attributes, Global Attributes, Event classifiers and Extensions.
There are different types of primitive attributes. The \emph{String} and \emph{Date} attributes are stored as JSON strings. JSON numbers represent \emph{floats} and \emph{integers}. \emph{Boolean} values are stored as JSON Boolean. The \emph{List} values is represented as an array of JSON Objects. Lastly, the \emph{Container} is stored as JSON object.
A \emph{log} object contains 0 or more \emph{trace} objects, which are stored in the \emph{traces} array of JSON objects. Each trace describes the execution of one specific instance, or case, of the logged process. Every trace contains an arbitrary number of \emph{events} objects. In addition every \emph{trace} has its own attributes stored in the \emph{attrs} object. Below is an example representation of the basic structure

\begin{verbatim}
{ 
  "traces": [{
    "attrs":{
        "name":"Mohamed",
        "age": "19",
    },
    "events": [{
        "concept:name": "Activity 1",
        "date": "2013-10-21T13:28:06.419Z",
        "org:resource":"Bob"
        },
        {
        "concept:name": "Activity 2",
        "date": "2013-10-21T13:28:06.419Z",
        "org:resource":"Alice"
        }]
    }]
}
\end{verbatim}

Because the string has the same power as the \emph{ID}, the \emph{ID} data type is not supported in JXES. Below is an example for the representation of every attribute-type.

\begin{verbatim}
{
    "string": "hi",
    "date": "2013-10-21T13:28:06.419Z",
    "int":1,
    "float": 1.0,
    "boolean": true,
    "list": [{"key":1},{"key":2},
        {"new key":"new value"}],
    "container":
        {
            "key":1,
            "key-2":"value 2"
            }
}
\end{verbatim}
To represent \emph{nested attributes} in JXES the \emph{container} with two keys with the names \emph{value} and \emph{nested-attrs} is reserved. 
Which means that every \emph{container} with any of the keys \emph{value} or \emph{nested-attrs} is reserved by JXES and can not be used by the user.
Every other container is allowed.

Below is an example for a nested attribute container.

\begin{verbatim}
{
   "Person":
    {
      "value":1,
      "nested-attrs":{"name":"Mohamed",
                    "age":19,
                    "married":false}
    }
}
\end{verbatim}

\emph{Global attributes} are attributes that are understood to be available and properly defined for each element on their respective level throughout the document.
The log object holds two lists of global attributes for the \emph{trace} level and for the \emph{event} level. This means, a global attribute on the event level must be available for every event in every trace.
In JXES, we have defined elements for the \emph{trace} level and the \emph{event} level. They are both stored in under the element \emph{global-attrs} as nested elements. Below is an example for a global attribute.
\begin{verbatim}
{
   "global-attrs":{"trace":{"Key 1":1},
                "event":{"Key 2":2}}
}
\end{verbatim}

\emph{Event Classifiers}  assigns to each event an identity, which makes it comparable to other events (via their assigned identity). The JXES format makes event classification configurable and flexible, by introducing the concept of
event classifiers. The classifier name is stored in the key and the classifier keys are stored as an array of strings.
An example of event classifiers can be found below.
\begin{verbatim}
{
  "classifiers":{
    "Activity classifier":
    ["concept:name","lifecycle:transition"]
  }
}
\end{verbatim}

\emph{Extensions} is a set of
attributes on any levels of the XES log hierarchy (log, trace, event, and meta for nested attributes). 
Extensions have many possible uses. One important use is to introduce a set of commonly
understood attributes which are vital for a specific perspective or dimension of event log analysis (and which may even not have been foreseen at the time of designing the XES standard).
They are stored in an array.
Every object in this array represents an extension.
The \emph{name}, \emph{concept} and \emph{uri} are stored as key value pairs.
See the following example of extension definitions.

\begin{verbatim}
{
   "extensions":[{"name":"Test",
       "prefix":"concept",
       "uri":"http://www.test.org/test.xes"
       }]
}
\end{verbatim}

\begin{figure}[htb!]
\includegraphics[width=0.5\textwidth]{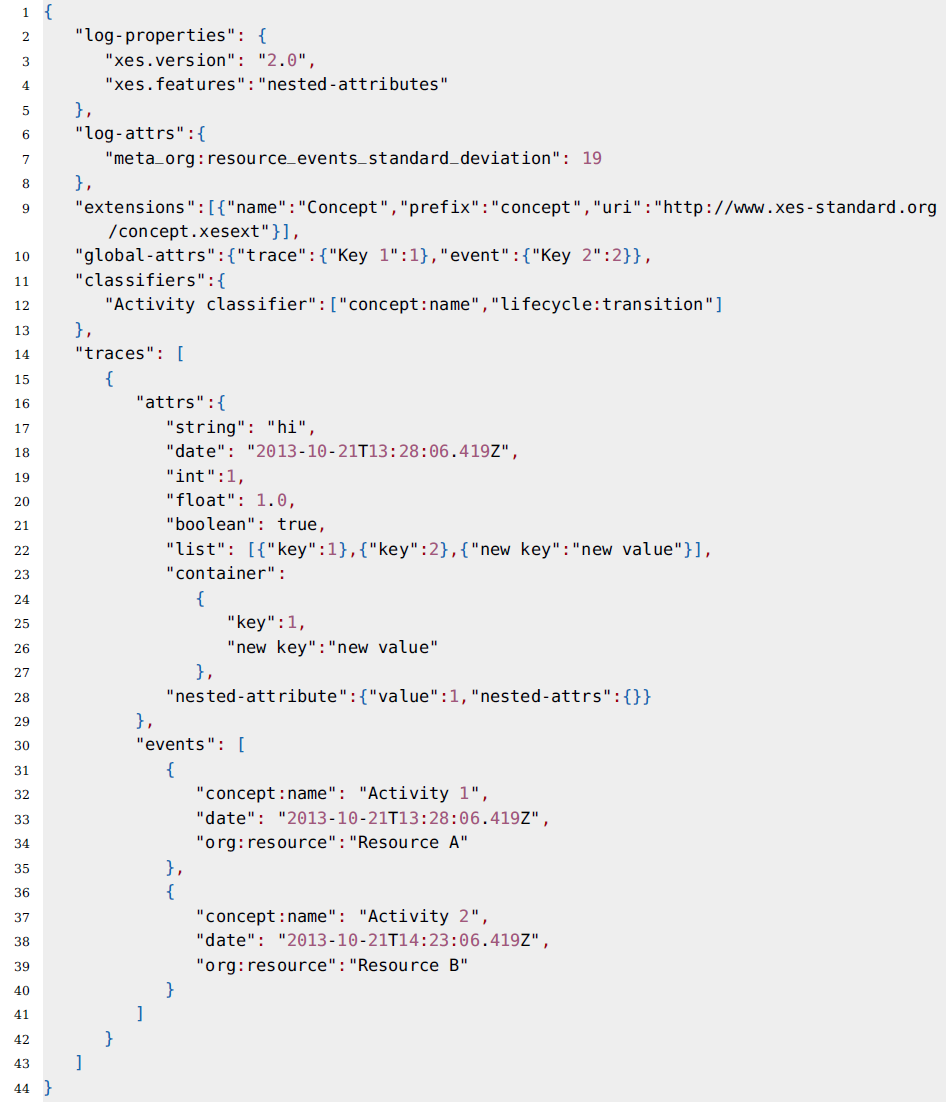}
\vspace{-0.5cm}
\caption{Full JXES template.}
\label{fig:xes}
\end{figure}

Figure~\ref{fig:xes} shows the full format of the JXES event log. 

\section{Implementation}
\begin{figure}[htb!]
\includegraphics[width=0.5\textwidth]{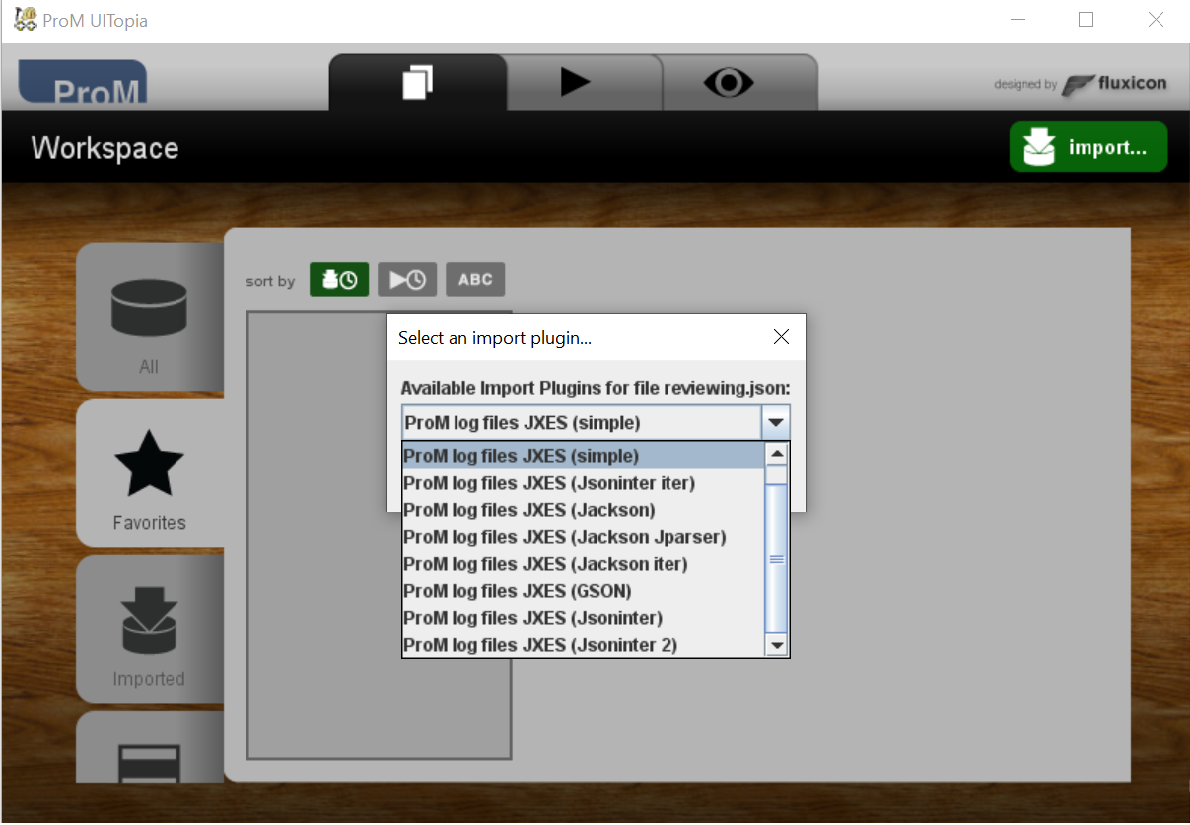}
\caption{Available JSON import parsers for JXES event log.}
\label{fig:importoptions}
\end{figure}

\begin{figure}[htb!]
\includegraphics[width=0.5\textwidth]{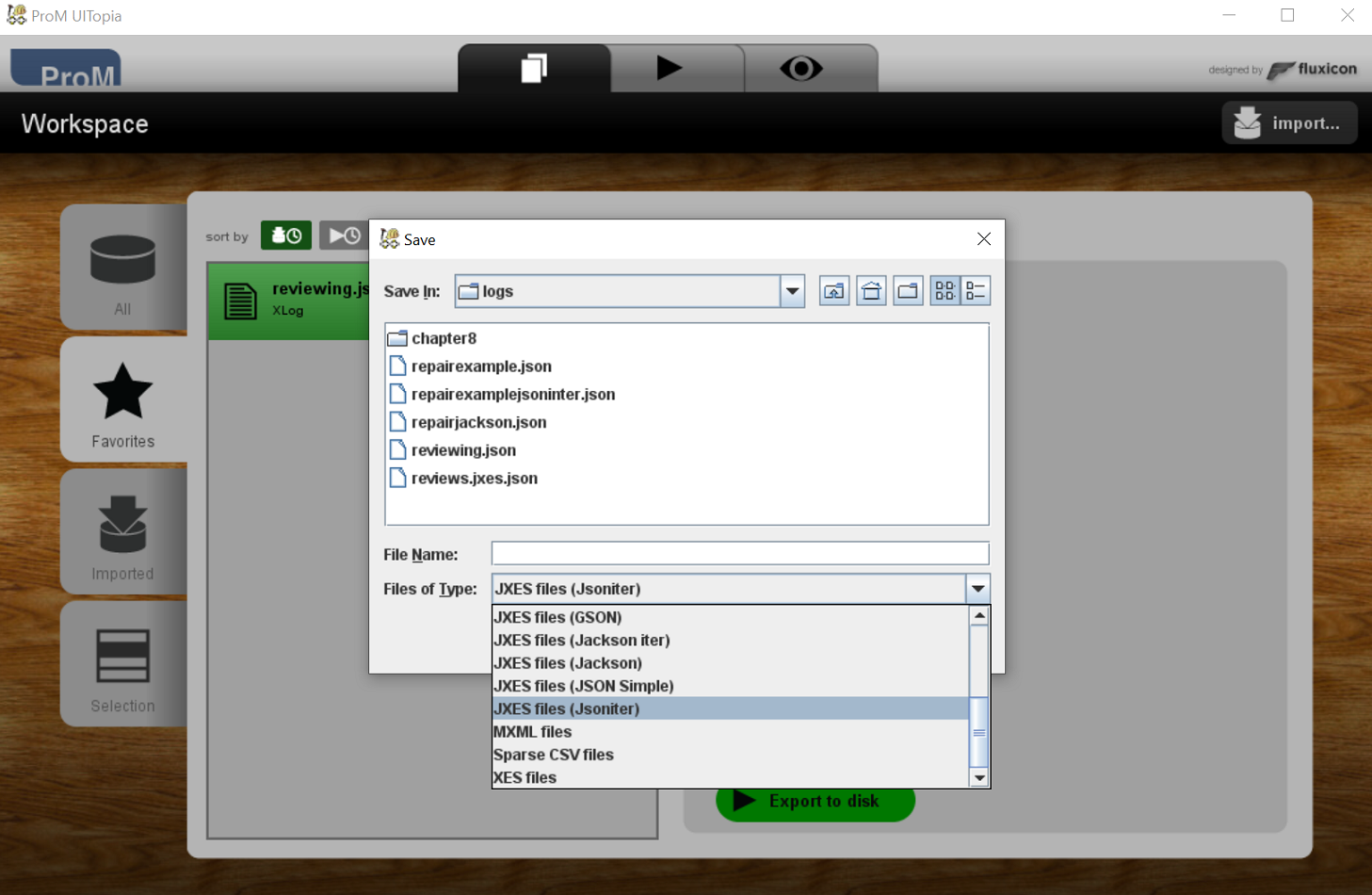}
\caption{Available JSON export parsers for JXES event log.}
\label{fig:export_options}
\end{figure}

The basic idea is to enable usage of JSON format of event logs in the ProM tool. To achieve this we did a market research of the top and best performing Java JSON parsers.
The plugin to import and export the JSON file consists of 4 different parser implementations of import as well as export. The parsers that have been implemented are Jackson, Jsoninter, GSON and simple JSON.
When the user clicks on "Import" in ProM tool and chooses a JSON file, the import options related to JXES are displayed where the user can choose from one of the parsers to import as shown in Figure~\ref{fig:importoptions}.
When the user clicks on "Export to disk" option, the console to save is displayed with Export parser options as shown in Figure~\ref{fig:export_options}.

\section{Evaluation}
We used 2 real-life event logs and 2 artificial event logs for our analysis. Table \ref{tab:logchar} highlights some of the key characteristics of the real-life event logs. The Level D2 log contains all standard attribute extensions like lifecycle, cost, concept and time extensions of an event log. The Flag X2 log is a Level D2 extended with attributes from non-standard XES extensions and/or attributes without an extension.

The evaluation of parsers was done alongside to the existing XES Naive and XES zipped implementations in ProM. The machine used to run the evaluation is equipped with a 4-core Intel i7 processor and 8 GB of RAM..

The criteria considered for evaluation are (1) speed (2) memory usage and (3) size.

To test the speed and memory of the different parsers, the file is imported/exported thrice and the average of the 3 runs is recorded in the tables. 
\begin{table}
\caption{Log characteristics.}
\label{tab:logchar}
\resizebox{\columnwidth}{!}{
\begin{tabular}{|l|*{6}{c|}}\hline

\makebox{}
& \makebox{traces}&\makebox{events}&\makebox{variants}
&\makebox{distinct activities} &\makebox{max trace length} \\\hline\hline
BPIC15\_5 & 1156 & 59083 & 1153 & 389 & 154 \\\hline
BPIC17 & 31509 & 1202267 & 4047 & 26 & 180 \\  \hline
Level D2 & 1104 & 11855 & 60 & 8 & 24 \\  \hline
Flag X2 & 1104 & 11855 & 60 & 8 & 24 \\  \hline
\end{tabular}
}

\end{table}
\subsubsection{Speed}
The best result for JSON parsers is highlighted in green. The unit of time specified in Table \ref{tab:is} and Table \ref{tab:es} is in milliseconds.

\begin{table}
\caption{Import Speed.}
\label{tab:is}
\resizebox{\columnwidth}{!}{
\begin{tabular}{|l|*{6}{c|}}\hline

\makebox{}
& \makebox{Simple JSON}&\makebox{GSON}&\makebox{Jackson}
&\makebox{Jsoninter} &\makebox{XES Naive}  &\makebox{XES zipped} \\\hline\hline
BPIC15\_5 & 1644.33 & \colorbox{cyan}{1342} & 1861 & 62269 & 1950 & 1866.7 \\\hline
BPIC17 & 50080.2 & 65741.66 & \colorbox{cyan}{29326.33} & ERROR & 24248.67 & 17476 \\  \hline
Level D2 & \colorbox{cyan}{293.33} & 333.33  & 317.33 & 367 & 462.67 & 259 \\  \hline
Flag X2 & 444.33 & 390.33.67 & 421.66 & \colorbox{cyan}{326} & 248 & 268 \\\hline
\end{tabular}
}
\end{table}

\begin{table}
\caption{Export Speed.}
\label{tab:es}

\resizebox{\columnwidth}{!}{

\begin{tabular}{|l|*{6}{c|}}\hline
\makebox{}
& \makebox{Simple JSON}&\makebox{GSON}&\makebox{Jackson}
&\makebox{Jsoninter} &\makebox{XES Naive}  &\makebox{XES zipped} \\\hline\hline
BPIC15\_5 & 874 & 679 & 707 & \colorbox{cyan}{477} & 1915.33 & 1797 \\\hline
BPIC17 & Error & 6336.33 & 9432.66  & \colorbox{cyan}{2925} & 20726  & 33433 \\\hline
Level D2 & 180.33 & 164.33 & 68 & \colorbox{cyan}{66.33} & 165 & 1750 \\\hline
Flag X2 & 214.33 & 205 & 96.66  & \colorbox{cyan}{73.33}  & 260.33 & 537 \\\hline
\end{tabular}
}
\end{table}

\subsubsection{Memory}
The memory consumption for different parsers and the XES implementation is computed using difference of the JAVA Runtime memory methods \emph{totalMemory} and \emph{freeMemory}.
It is noticeable that the Jackson JSON parser uses significantly less memory than the other parsers. And also that Simple JSON uses little less memory when it comes to average sized files. 
The unit of memory given in Table \ref{tab:imc} and Table \ref{tab:emc} is MBs.

\begin{table}
\caption{Import Memory Consumption.}
\label{tab:imc}

\resizebox{\columnwidth}{!}{

\begin{tabular}{|l|*{6}{c|}}\hline
\makebox{}
& \makebox{Simple JSON}&\makebox{GSON}&\makebox{Jackson}
&\makebox{Jsoninter} &\makebox{XES Naive}  &\makebox{XES zipped} \\\hline\hline
BPIC15\_5 &  1534.08  & 1374.84  & \colorbox{cyan}{953.878} &  2821.53  &  1421.53  & 1440.92 \\\hline
BPIC17 & 2838.03 & 3060.57 & \colorbox{cyan}{2726.59}  & ERROR & 2264.38  & 1329.12 \\\hline
Level D2 & 1168.21 & \colorbox{cyan}{572.93}  & 378.79 & 576.80  & 1324.92  & 686.00 \\\hline
Flag X2 & 517.17 & 630.73 & \colorbox{cyan}{284.85} & 754.37 & 356.15 & 780.43 \\\hline
\end{tabular}
}
\end{table}

\begin{table}
\caption{Export Memory Consumption.}
\label{tab:emc}

\resizebox{\columnwidth}{!}{

\begin{tabular}{|l|*{6}{c|}}\hline
\makebox{}
& \makebox{Simple JSON}&\makebox{GSON}&\makebox{Jackson}
&\makebox{Jsoninter} &\makebox{XES Naive}  &\makebox{XES zipped} \\\hline\hline
BPIC15\_5  & \colorbox{cyan}{448.55} & 679.3 & 869.53 & 1040.04 & 355.528 & 1708.61 \\\hline
BPIC17  & ERROR & 1441.65  & \colorbox{cyan}{1293.80} & 2255.95 &  1958.743  & 1923.8 \\\hline
Level D2  & 461.12 & 279.97 & \colorbox{cyan}{96.80} & 301.9 & 710.00 & 659.80 \\\hline
Flag X2 & 467.20 & 264.58 & \colorbox{cyan}{110.27} & 302.96 & 350.29 & 305.34 \\\hline
\end{tabular}
}
\end{table}

It is clear from the results of the export plugin that JXES is up to 4x faster than XES. One reason is the reduced syntax that gets written when exporting JSON as compared to XML. 

It is also clear that the Jsoninter parser achieves better results than all others in terms of exporting speed. The speed improvement in Jsoninter can be attributed to its Dynamic Class Shadowing tri-tree feature.
It is noticeable that the Jackson JSON Parser uses significantly less memory than the other parsers. This performance can be attributed to the incremental parsing/generation feature which reads and writes JSON content as discrete events. And also that Simple JSON uses little less Memory when it comes to average sized files. 
In addition the Jsoninter JSON parsers uses the most memory in all cases.

\subsubsection{Size}
Table \ref{tab:lfs} provides the size in MBs for the files stored. The size improvement in JXES is obvious because the type is not specified and tag names are not written twice and the markup is not repeated.

We noted that there was no loss of information during the conversion from XES to JXES and vice versa. The only difference noted was the log version information in the header of XES file. We observed that the performance of the XES Naive importer is surprisingly good when compared with the JSON format. This can be attributed to the fact that XES format has the datatype specified in the tag whereas in JSON we need to parse it completely to determine the datatype.

\begin{table}
\centering
\caption{Log File Size.}
\label{tab:lfs}
\resizebox{0.9\columnwidth}{!}{

\begin{tabular}{|l|*{5}{c|}}\hline
\makebox{}
& \makebox{XES} &\makebox{JXES} &\makebox{XES zipped} \\\hline\hline
BPIC15\_5  & 43.9 & 29.6 & 1.6  \\\hline
BPIC17 & 552 & 323.7  & 28.2  \\\hline
Level D2 & 4.6 & 2.5 &  0.246  \\\hline
Flag X2   & 5.1 & 2.6 & 0.305  \\\hline

\end{tabular}
}
\end{table}

\section{ACCESS and DEMONSTRATION}
The JXES import and export plugins have been implemented with 4 different JSON parsers. The code is available in the SVN repository \cite{coderepo}. The sample logs for JXES format can be found under the tests/testfiles/ directory. The tutorial video of the parsers implemented as a ProM plugin is available at \url{https://youtu.be/sZ6UnTfSsFI}.

Figure~\ref{fig:importoptions} and Figure~\ref{fig:export_options} show the different parser options to import and export JSON event logs respectively. To run the import plugin, we will require the event log in JSON format shown in Figure~\ref{fig:xes}. 

The application can also be run by importing any of the event log in CSV or XES format and then converting it into the JSON format.

\section{CONCLUSION}
We have introduced JXES, which is a JSON format of event log which adheres to the XES principles. In this paper, we defined the JSON format of event log as defined by the IEEE XES standard. 
After defining the standard, we have provided 4 different implementations of import and export options for JSON event logs with different parsers.

Table \ref{tab:prs} shows the authors recommendation for parser choice in case of import and export considering speed and memory criteria.

It would be interesting to evaluate more  parsers (e.g., JSONP, fastJSON and ProtoBuf)

\begin{table}
\centering
\caption{Parser recommendation chart.}
\resizebox{0.75\columnwidth}{!}{

\begin{tabular}{|l|*{5}{c|}}\hline
\makebox{}
& \makebox{Speed} &\makebox{Memory} \\\hline\hline
Import  & Jackson & Jackson   \\\hline
Export & Jsoninter & Jackson    \\\hline
\end{tabular}
}
\label{tab:prs}
\end{table}

We hope that the JXES standard defined by this paper will be helpful and serve as a guideline for generating event logs in JSON format. We also hope that the JXES standard defined in this paper will be useful for many tools like Disco, Celonis, PM4Py, etc., to  enable support for JSON event logs.
We also hope that the plugin implemented in ProM will be useful for many with JSON event logs. 

\bibliographystyle{IEEEtran}
\bibliography{json}
\end{document}